# Bias current dependence of the spin lifetime in insulating $Al_{0.3}Ga_{0.7}As$

Jennifer Misuraca[1a], Joon-Il Kim[1], Jun Lu[2], Kangkang Meng[2], Lin Chen[2], Xuezhe Yu[2], Jianhua Zhao[2], Peng Xiong[1], Stephan von Molnár[1]

[1] *Department of Physics, Florida State University, Tallahassee, Florida, 32306, United States*
[2] *State Key Laboratory for Superlattices and Microstructrues, Institute of Semiconductors, Chinese Academy of Sciences, P.O. Box 912, Beijing 100083, China*

Abstract:

The spin lifetime and Hanle signal amplitude dependence on bias current has been investigated in insulating $Al_{0.3}Ga_{0.7}As$:Si using a three-terminal Hanle effect geometry. The amplitudes of the Hanle signals are much larger for forward bias than for reverse bias, although the spin lifetimes found are statistically equivalent. The spin resistance-area product shows a strong increase with bias current for reverse bias and small forward bias until 150 µA, beyond which a weak dependence is observed. The spin lifetimes diminish substantially with increasing bias current. The dependence of the spin accumulation and lifetime diminish only moderately with temperature from 5 K to 30 K.

---

[a] Author to whom correspondence should be addressed. Electronic mail: jmisuraca@bnl.gov.

The transport and accumulation of spins has been detected electrically in several semiconductors to date (GaAs [1,2], Si [3-6], InGaAs [7,8], Ge [9-12], AlGaAs[13]), and the bias and temperature dependences of the spin signals have been investigated in these materials. These studies are necessary in order to establish whether or not spin-dependent measurements in these materials can possibly lead to room temperature, nonvolatile, and low power applications for future information storage and logic processing technologies. An effective and frequently employed method for measuring spin accumulation and extracting spin lifetime is the three-terminal (3T) Hanle geometry [4,6]. The amplitude of the Hanle signal corresponds directly to the spin-splitting of the chemical potential (magnitude of the spin accumulation) and is expected to be linear with the bias current and independent of the current direction[14,15]. However, there is an observed asymmetry between forward and reverse biasing in most semiconductors. In n-doped GaAs, forward biasing of the injector (electrons extracted from the semiconductor into the ferromagnet) is the favorable, sometimes the only, configuration leading to a Hanle signal [1], while in p-type Si [4] and n-type Ge [12], the amplitudes of the reverse biased Hanle signals (carriers injected from the ferromagnet into the semiconductor) are found to be three and five times larger than those under the same forward biased currents, respectively. Moreover, recent experiments[4,16] which measured the spin accumulation in Si and Ge induced by tunneling injection of spin-polarized current showed a clear violation of the expected linear proportionality between the spin voltage and injection current. The Hanle signal sizes are also generally smaller as the temperature rises, regardless of the material [2,4,11]. The lifetime of the spin polarization is also an important parameter to research in order to make devices viable for spintronic applications. It must be long enough for information to be encoded and then decoded from a spin-polarized current. Spin lifetimes are shown to decrease with increasing temperature in GaAs [2] and n-type Ge [12], but stay relatively constant in heavily doped Si [6] and heavily doped p-type Ge [11].

Local and nonlocal spin accumulation has recently been detected electrically in the persistent photoconductor [17-19] $Al_{0.3}Ga_{0.7}As$:Si at various carrier densities by tuning the effective carrier density of the material in one sample *in situ* using photoexcitation[13]. The present letter focuses on the investigations of the bias and temperature dependences of the spin lifetime in this material at a carrier density of $n = 5 \times 10^{16}$ cm$^{-3}$, which is on the insulating side of the metal-insulator transition (MIT) ($n_c = 9 \times 10^{16}$ cm$^{-3}$, see Ref. 19), as well as on the source and drain asymmetry of the spin accumulation signals in a 3T Hanle effect geometry. Details of the Fe/AlGaAs heterostructure, device geometry, and fabrication procedure have been presented previously[13]. In this device, the Fe electrode that the spins accumulate under is 50 x 10 μm$^2$ and 5 nm thick[20]. The spin accumulation measurements were completed using the Hanle effect in a 3T geometry (see Fig. 1 (d) in Ref. 13 for a depiction of the measurement scheme). At this insulating carrier density, the amplitudes of the forward biased Hanle curves are found to be much larger than the reverse biased curves, with the amplitudes increasing as the bias current is increased in both instances. The spin lifetimes in this material are longest at the lowest forward bias measured and decrease significantly with increasing bias; they also decrease very moderately at higher temperatures.

In experiments where the ferromagnetic metal electrode is used as a source of electron spins, the Schottky junction is reverse biased, which means that the majority spins will be injected from the ferromagnet into the channel and will accumulate near the interface. In experiments where the same electrode is used as a drain for electron spins, the only difference is the direction of the current; in this case, the Schottky barrier is forward biased and the unpolarized majority spins from the AlGaAs channel will be preferentially transmitted into the spin polarized states in the ferromagnet, leaving the minority spins accumulating in the channel near the interface. In both cases, a small perpendicular magnetic field is swept and will cause these accumulated spins to precess and dephase, which is referred to as the Hanle effect.

The Hanle signal is determined by symmetrizing the raw data to remove the anti-symmetric background, such as from local Hall effects, and then by subtracting an even polynomial, which is best-fit to the high-field data beyond the Hanle signal, to eliminate the remaining background. These data can then be modeled using a Lorentzian fitting procedure which is determined by the following equation:

$$L_z(B) = \frac{L_0}{1 + \left(\frac{g\mu_B B \tau_s}{\hbar}\right)^2} \tag{1}$$

where $L_0$ is the amplitude of the curve, $g$ is the electron g-factor, $\mu_B$ is the Bohr magneton, $B$ is the magnetic field, $\tau_s$ is the spin lifetime, and $\hbar$ is the reduced Planck's constant [21]. The spin lifetime can be extracted since it will be inversely proportional to the full width at half maximum (FWHM) of the curve (determined by solving for $B$ if $L_z$ is equal to half of $L_0$). It is generally agreed that this method provides a lower bound for the value of the spin lifetime [4] and it is frequently implemented in the literature in electrical spin accumulation experiments utilizing various semiconducting materials [4-6,11,22]. Using the 1D spin drift-diffusion model[2,13] instead to fit the data leads to an increase of approximately a factor of two for all of the lifetimes measured; here we report the lower bound values.

The bias dependence of the Hanle signal at 5 K is shown in Fig. 1 for forward and reverse biasing. The Hanle data are shown as filled circles and the solid lines are the Lorentzian fits to the data. As one can see, the Lorentzian does not fit well at high biases around zero field. This sharp increase at very small fields is due to dynamic nuclear polarization, which is caused by the momentum transfer from the electrons to the nuclei via the hyperfine interaction during the spin injection[23]. It is commonly detected in the spin transport and accumulation of GaAs [24,25]. The voltage drop across the junction is plotted as a function of current density in the inset of Fig. 1. One can see that there is also an asymmetry in forward and reverse bias in this IV curve, qualitatively similar to what is seen in GaAs[26].

The amplitudes of the Hanle signals are plotted (in red) as a function of bias current in Fig. 2. The amplitude increases with increasing forward bias as expected, and the amplitudes of the reverse biased curves do not depend as strongly on the bias current. This discrepancy is more noticeable in the inset of Fig. 3, where low bias curves from Fig. 1 (at I = 50 μA and 100 μA) are enlarged so that the signal size of reverse and forward biased curves can be compared. This discrepancy continues to increase as the bias is increased.

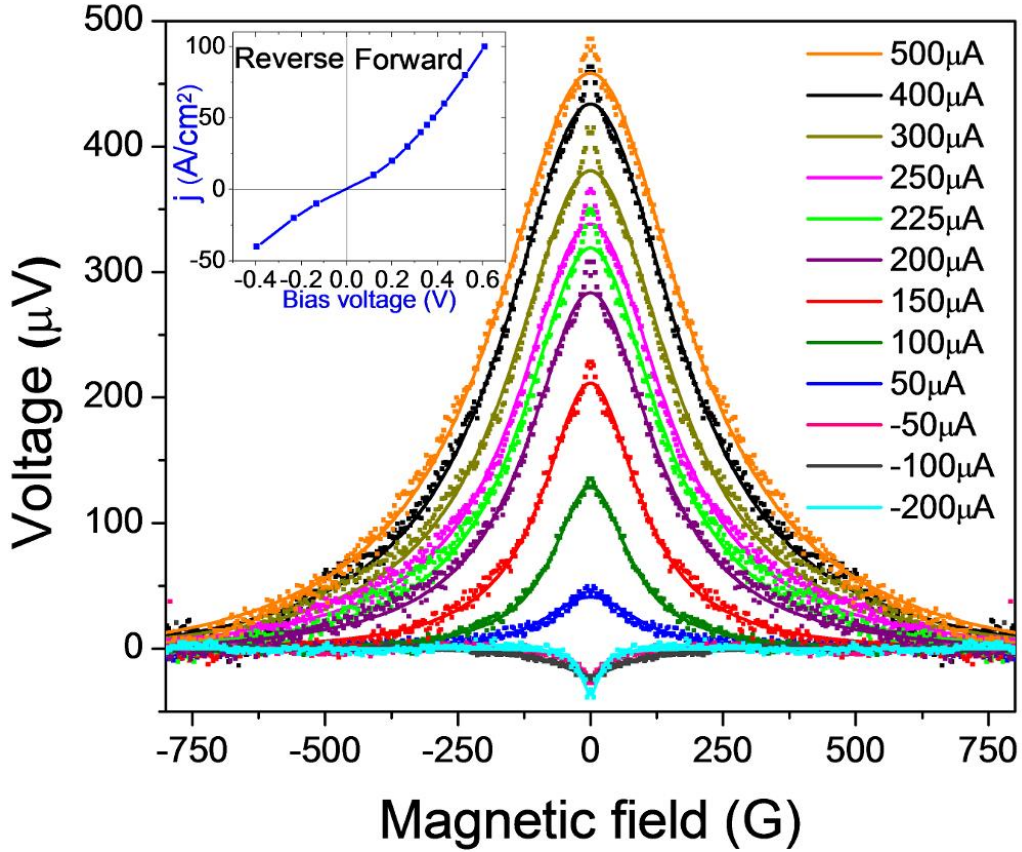

**Figure 1:** Bias current dependence of spin accumulation at 5 K with electrode as source (negative bias) and drain (positive bias). Inset: Current density versus bias voltage of the junction.

The mechanism behind the strong bias dependence in 3T spin accumulation measurements is still not completely understood. In a simple picture, one would expect a linear response under both forward and reverse bias, though this is almost never the case experimentally[1,4,12,16]. For devices with thin Schottky barrier injectors and detectors, there is pronounced asymmetry between the spin accumulation signals under forward and reverse bias. A systematic study by Hu et al.[26] revealed a correlation between spin accumulation at such a junction and the formation of a quantum well-like accumulation layer in the barrier: The filling of the quantum well states is different under forward and reverse bias, resulting in the difference in spin accumulation[26]. This effect is particularly strong for a thickness of 15 nm for the $n^+$ doped layer, at the onset of tunneling[26]. The strong asymmetry of spin accumulation magnitude with respect to the bias direction in Fig. 1 is consistent with this picture and supports the interpretation that the signals are due to spin accumulation.

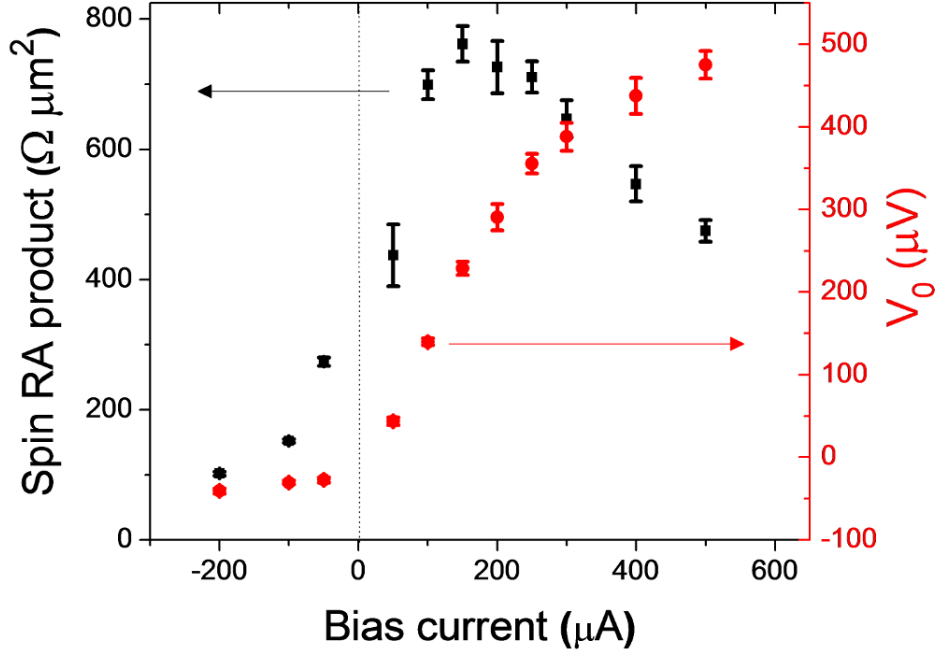

**Figure 2: Dependence of the Hanle signal amplitude and spin-RA product on bias current**.

The experimentally determined spin resistance-area product (spin-RA product) is defined as:

$$R_s A = \frac{\Delta V_{3T}(B_z = 0)}{I} A, \qquad (2)$$

where $\Delta V_{3T}(B_z = 0)$ is the amplitude of the 3T Hanle signal at zero field, $I$ is the bias current applied, and $A$ is the area of the contact that the spins accumulate under. A constant spin-RA product versus bias would indicate that the Hanle signal amplitude is linear in bias current [4], but it is not linear in the present experiment (see Fig. 2 in black) or in a previously reported work in Si at 5 K [4]. In that reference, the spin-RA product varies with bias asymmetrically, weakly when spins are being injected into and strongly when spins are being extracted from the semiconductor. However, in the present experiment, the spin-RA product shows a strong dependence on bias for injection and extraction at low biases. It reaches a maximum at a forward bias of I = +150 µA, after which it exhibits a weaker decrease with further increase of the forward bias. A calculation by Valenzuela et al. [27] shows a comparable bias dependence of the dynamic polarization of spin accumulation in Al in contact with a ferromagnet, including a maximum at a nonzero value at reverse bias. It should be noted that both the junction RA product ($R_j A$) and the spin resistance of the channel ($\rho_{ch}\lambda_s$, where $\lambda_s$ is the spin diffusion length) do not show this maximum under forward bias. The junction RA product (spin resistance of the channel) decreases from 1 x $10^6$ Ω-µm$^2$ (1000 Ω-µm$^2$) at 50 µA to 6 x $10^5$ Ω-µm$^2$ (330 Ω-µm$^2$) at 500 µA.

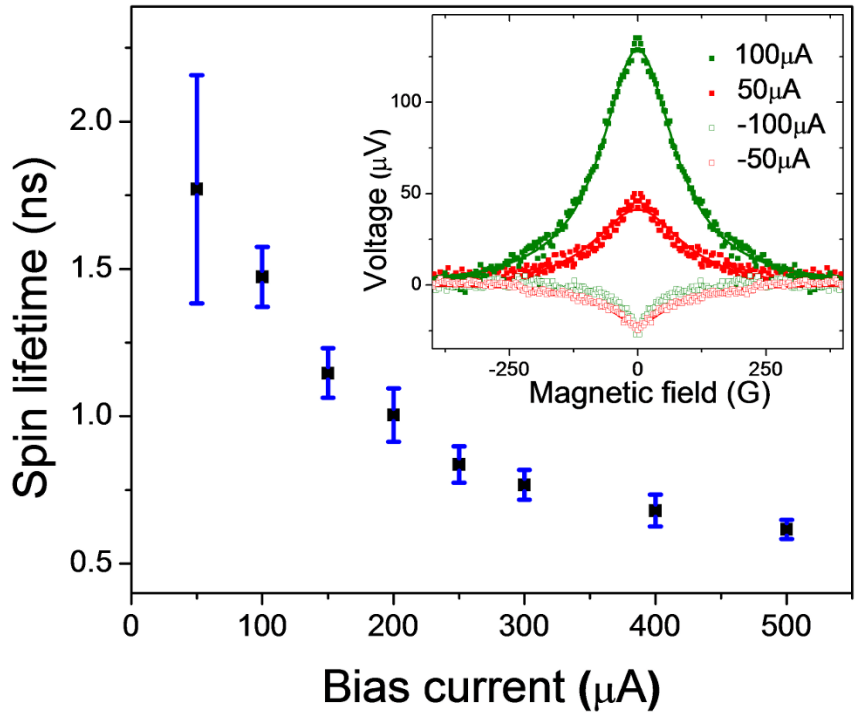

**Figure 3: Dependence of spin lifetime on bias current. Inset: Comparison of Hanle signal for forward and reverse bias configurations.**

Fig. 3 shows the spin lifetime dependence on bias current for forward biased curves. The lifetimes of the reverse biased curves are not plotted here because the large discrepancies in repeated measurements lead to much larger errors than for the forward biased measurements; these values and associated errors can be found in Table 1. The large discrepancies arise from the decreased signal to noise ratio of the curves for spin injection (reverse bias) due to the much weaker increase of the Hanle amplitude with bias current than in the case of spin extraction (forward bias). Even though the Hanle signal amplitudes show a large asymmetry in the forward versus reversed bias data as previously discussed, the spin lifetimes obtained from the FWHM of the Lorentzian fits of each bias are the same within the precision of the experiment, as shown in Table 1. The value of the spin lifetime obtained in this 3T experiment can be compared to non-local 4T transport results published previously on the same device[13]. For this carrier density and a bias of $265 \pm 8$ µA, the spin lifetime obtained from a 4T transport measurement is $1.78 \pm 0.33$ ns using the same Lorentzian fitting procedure. If we compare that to the spin lifetime obtained using the 3T spin accumulation method at comparable bias (from Fig. 3), the spin lifetime obtained is ~ 0.85 ns, within a factor of 2. This gives additional credence to the idea that these 3T spin accumulation measurements are indeed measuring a spin signal.

**Table I: Comparison of spin lifetime $\tau_s$ and Hanle signal amplitude $V_0$ for source and drain.**

|        | I (µA) | $\tau_s$ (ns)   | $V_0$ (µV)          |
|--------|--------|-----------------|---------------------|
| Drain  | 50     | $1.77 \pm 0.39$ | $43.73 \pm 4.67$    |
| Source | -50    | $1.59 \pm 0.31$ | $-27.44 \pm 3.25$   |
| Drain  | 100    | $1.47 \pm 0.10$ | $139.84 \pm 4.17$   |
| Source | -100   | $2.34 \pm 0.78$ | $-30.48 \pm 2.30$   |

The maximum spin lifetime for the forward biased curves occurs at the lowest current measured. However, these low current measurements are limited by noise, as can be seen from the larger error associated with the 50 µA datum point in Fig. 3. For GaAs at low temperatures, a small bias electric field over a spin transport channel left the spin lifetime unaffected[28]. Thus, the transport measurements could be completed without negatively affecting the spin lifetime at any bias below this threshold. In the present data, we see an immediate substantial decrease in the spin lifetime with bias most likely due to the large junction resistance and necessary injection current for a measurable spin accumulation using the 3T Hanle geometry as opposed to a 4T spin transport geometry.

Finally, the temperature dependence of the Hanle signals from spin accumulation has been investigated in insulating AlGaAs using the 3T configuration (see Fig. 4). The noise in the measurements corresponds to a very small instability in the temperature control and the data at 25 K and 30 K show a few points at zero bias which fall below the Lorentzian fit. The authors believe this is an artifact of the symmetrization which only appears due to the low signal to noise ratio. The spin lifetime in GaAs from Ref. 2 decreases from 24 ns at 10 K to 4 ns at 70 K. In the present work, we find a much less pronounced temperature dependence (shown in the inset of Fig. 4) where the spin lifetime varies from 1 ns to 700 ps for insulating AlGaAs from 5 K to 30 K. The smaller values for the lifetime are expected since there is much more disorder in the highly Si-doped AlGaAs samples used in this study than in the GaAs samples doped to around the MIT in Ref. 2. Also, the moderate temperature dependence of the spin lifetime in AlGaAs is not out of the ordinary; for example, in Refs. 6 and 11, the spin lifetime of highly doped Si and p-type Ge, respectively, show very little change with temperature.

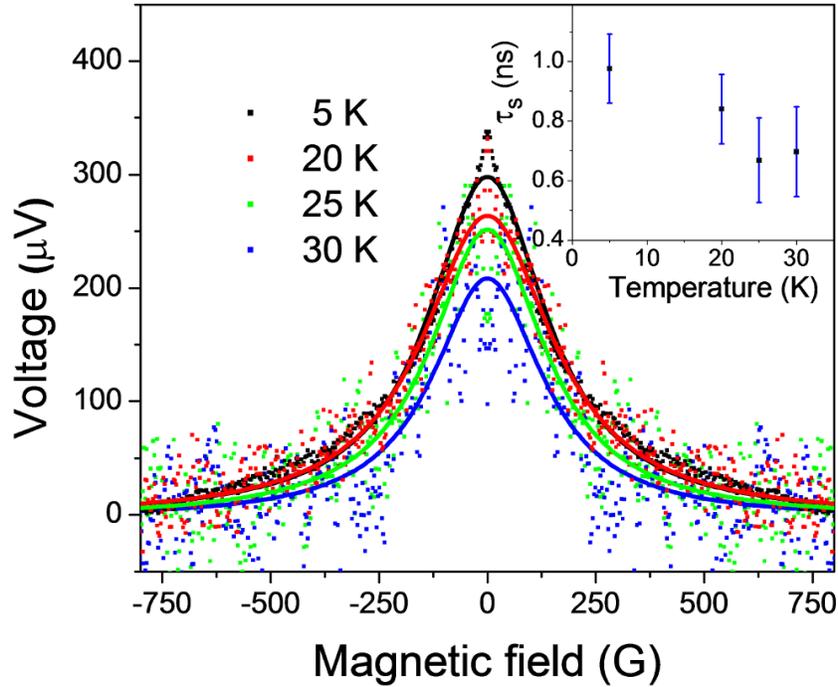

**Figure 4: Temperature dependence of the Hanle signal (forward biased) with I = 200 µA. Inset: Spin lifetime versus temperature.**

In conclusion, the spin lifetime dependence on bias and temperature has been investigated in insulating AlGaAs.  These dependences are imperative for elucidating the spin lifetimes in upcoming, more detailed experiments using AlGaAs.  At this carrier density, the amplitudes of the Hanle signals were much larger for forward biasing, as is the case in slightly metallic GaAs.  The spin-RA product was found to have a strong dependence on bias from negative values (spin injection) until a maximum at a forward bias of 150 µA.  The spin lifetimes were found to diminish substantially with increased bias, and diminish only moderately with increasing temperature, which is less pronounced than in GaAs.  Knowing that the spin lifetime is in the nanosecond range at low bias and that it is only moderately affected by temperature makes this an interesting material to study which may be well-suited for technological applications.  At the very least, the ability to tune the carrier density in this material *in situ* makes it an ideal material to study how the doping levels in semiconductors affect spin-dependent properties, which can lead to a greater understanding of how to optimize the parameters of semiconducting materials which can be utilized for spintronic devices.

This work has been supported by the NSF under Grants DMR-0908625 and DMR 1308613 and the NSFC under Grant 11228408.